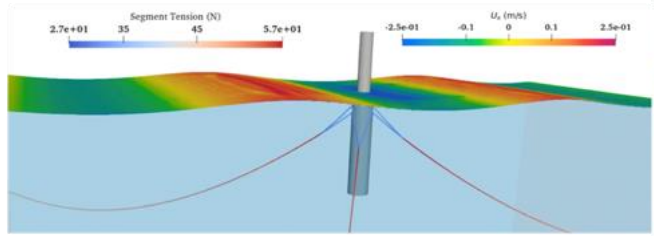
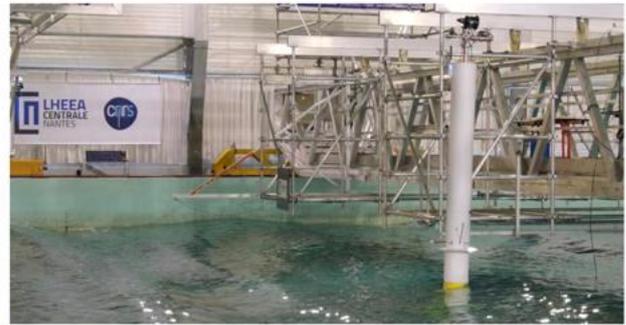

19e Journées de l'Hydrodynamique
26 – 28 novembre 2024, Nantes

# CONCEPTION, CONSTRUCTION ET MISE AU POINT D'UN NOUVEAU BASSIN DE CARENES A L'UNIVERSITE DE SOUTHAMPTON : UNE DECENNIE D'EFFORTS

# DESIGN, DEVELOPMENT AND COMMISSIONING OF A NEW TOWING TANK AT THE UNIVERSITY OF SOUTHAMPTON: A DECADE OF ENDEAVOUR

**B. MALAS[1,2]**
bertrand.malas@ec-nantes.fr

[1] Laboratoire de recherche en Hydrodynamique, Énergétique et Environnement Atmosphérique, Ecole Centrale de Nantes, 1 rue de la Noë, 44321 Nantes, France
[2] Faculty of Engineering and Physical Sciences, University of Southampton, Building 185, Boldrewood Campus, Burgess Road, Southampton SO16 7QF, United Kingdom

**Résumé**

L'université de Southampton au Royaume-Uni est réputée mondialement dans le domaine de l'architecture navale. Pendant plusieurs décennies, les chercheurs en hydrodynamique ont discuté la construction d'un bassin de carènes de taille moyenne, afin de pouvoir accueillir les activités d'enseignement et de recherche ainsi que les prestations commerciales auparavant délocalisées. La rénovation du campus de Boldrewood, débutée au début des années 2010, a ouvert la porte à la construction de ce bassin de 138 m de long, 6 m de large et 3,5 m de profondeur. Mis en service début 2022, il est équipé d'un générateur de houle composé de 12 panneaux indépendants et d'un chariot innovant permettant d'atteindre une vitesse de 10 m/s. La conception, construction et mise au point de ce moyen d'essais ont été jalonnées de nombreux obstacles et retards qui sont décrits ici. Les solutions techniques et équipements retenus sont détaillés. Les essais de validation sont également présentés.

**Summary**

The University of Southampton in the United Kingdom is worldly renowned in the naval architecture domain. For several decades, the researchers in hydrodynamics have been discussing the need of a middle-sized towing tank to accommodate the teaching, research and commercial experiments that previously had to be outsourced. In the early 2010s, the renovation of the Boldrewood campus was decided and opened the way for the construction of this 138 m long, 6 m wide and 3.5 m deep facility. Commissioned in 2022, the tank is equipped with a 12 independent paddles wavemaker and an innovative carriage capable of speeds of up to 10 m/s. The design, construction and commissioning of the experimental facility were subject to many hurdles and delays that are described here. Details about the technical solutions and chosen equipment are provided. Validation experiments are also described.

**Préambule :**

Cet article est une adaptation et traduction d'un article publié en janvier 2024 dans la revue « International Journal of Marine Engineering », éditée par le RINA (Royal Institution of Naval Architects, référence [1]).

# I - Introduction

Le Royaume-Uni a historiquement toujours été à l'avant-garde de l'hydrodynamique expérimentale, de la naissance du premier bassin de carènes construit en 1870 à Torquay par William Froude à la seconde moitié du XXème siècle quand plus de dix bassins de carènes et océaniques étaient en opération dans le pays (référence [2]), dans les domaines militaires, de la recherche ou commerciaux.

L'université de Southampton, réputée mondialement pour son enseignement et sa recherche en matière d'architecture navale a, après plusieurs décennies passées à utiliser des moyens d'essais extérieurs, décidé au début des années 2010 de construire un bassin de carènes de moyenne taille.

Cet article documente les processus de conception, construction et mise au point de ce nouveau bassin ainsi que les problèmes rencontrés et les retards subis. Les équipements installés dans le bassin sont également présentés, ainsi que les essais de validation.

# II - Contexte et historique

L'université de Southampton est située dans la ville du même nom sur la côte sud de l'Angleterre. Créée en 1862, elle est membre fondatrice du Russel Group (réseau de vingt-quatre des plus grandes universités au Royaume-Uni axées sur la recherche scientifique)[1]. Lors de l'année scolaire 2022-2023, elle accueillait environ 26000[2] étudiants et employait environ 6200 personnes[3].

## II - 1 Le besoin

Le département d'architecture navale de l'université de Southampton (Ship Science) a été créé dans les années 60, afin de regrouper les activités de recherche déjà existantes et de lancer un programme d'enseignement dans ce domaine. Quelques années plus tard, WUMTIA (Wolfson Unit for Marine Technology and Industrial Aerodynamics), une petite unité d'ingénieurs sans contraintes de recherche ou d'enseignement, était créée afin de commercialiser le savoir généré au sein de Ship Science (référence [3]).

Si l'université possédait depuis longtemps deux souffleries de bonne taille, le seul bassin de carènes disponible sur le campus, le Lamont tank, ne faisait que 30 m de long, 2,4 m de large et 1,2 m de profondeur, et était logiquement limité à l'enseignement. Les activités expérimentales de recherche (références [4] à [7]) et commerciales étaient donc délocalisées dans les moyens d'essais environnants, tels que le bassin de GKN Aerospace sur l'île de Wight (200 m de long - fermé en 2008), le bassin de QinetiQ à Haslar (270 m de long) ou le bassin de Solent University à Southampton (60 m de long), avec tous les impératifs liés aux déplacements et les risques techniques associés.

Il faut également noter que depuis les réformes de la fin des années 90 et l'introduction des frais d'inscription à l'université au Royaume-Uni puis leur augmentation progressive depuis lors (l'année d'études en génie maritime à Southampton coûte en 2024/2025 £9250 pour les étudiants anglais et £28800 pour les étrangers[4]), il y a une intense compétition entre universités pour le recrutement d'élèves. La présence d'infrastructures de recherche de moyenne ou grande taille accessibles aux étudiants est un des facteurs pour attirer les futurs élèves.

---

[1] https://www.southampton.ac.uk/about/reputation/history.page
[2] https://www.hesa.ac.uk/data-and-analysis/students/where-study#provider
[3] https://www.hesa.ac.uk/data-and-analysis/staff/working-in-he
[4] https://www.southampton.ac.uk/index.php/courses/maritime-engineering-degree-meng#fees

II - 2 Le site

A partir des années 90, plusieurs sites possibles dans et autour de Southampton ont été étudiés pour la construction de ce nouveau bassin, et notamment le campus de Boldrewood, situé à proximité du campus principal de Highfield. Construit en 1975 pour la faculté de médecine, ses bâtiments étaient au début des années 2000 très dégradés et coûteux pour l'université. En 2006, il fut décidé de relocaliser les activités du campus à l'hôpital et de le démolir et reconstruire en le consacrant cette fois à la faculté d'ingénierie. En 2010, le campus fermait et les travaux commençaient.

II - 3 Historique

Si le bassin de carènes était inclus dès le départ dans le plan du nouveau campus, il n'était pas complètement financé jusqu'en 2012, quand les coûts de construction ont soudainement baissé au Royaume-Uni, suite à la crise financière mondiale de 2007-2008. Cette baisse a permis de facto le lancement de la construction du bassin plus tôt que prévu. Une fois la décision de la construction prise, les dimensions principales du bassin et donc du bâtiment ont été rapidement fixées. La longueur du bassin a été déterminée par les contraintes du campus : en effet, une conduite de gaz et un arbre protégé par le plan local d'urbanisme la fixaient à 138 m. La largeur de 6 m est assez conventionnelle pour un bassin de cette taille et la profondeur a été limitée à 3,5 m par les coûts d'excavation.

La vitesse du chariot a été spécifiée à 12 m/s dans les deux directions, ce qui était très ambitieux pour un bassin de cette longueur.

Enfin, la section transversale du bassin a été rapidement fixée, ainsi que le plan du bâtiment construit autour du bassin, mais aussi des autres laboratoires hébergés à l'intérieur. Les fondations consistent en 340 piliers en acier de 400 mm de diamètre, enfoncés à une profondeur de 21 m dans le sous-sol argileux (Figure 1).

Ces premières spécifications ont permis de lancer la construction mi-2013 (Figure 2), assurée par l'entreprise Wates[5] avec le soutien des sociétés pcdm[6] et BuroFour[7]. Il était alors estimé que le bassin de carènes serait complètement opérationnel en 2016, voire en 2017.

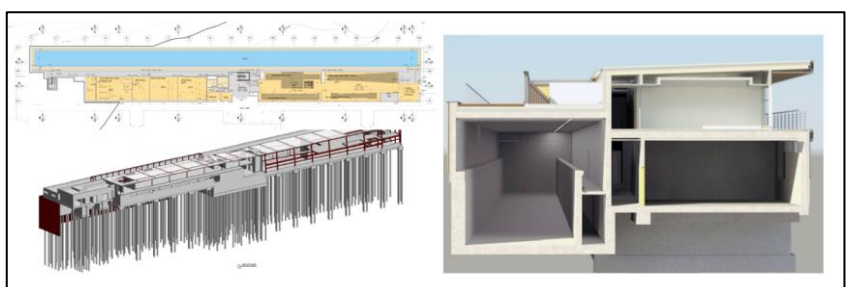

Figure 1 : Caractéristiques du bâtiment

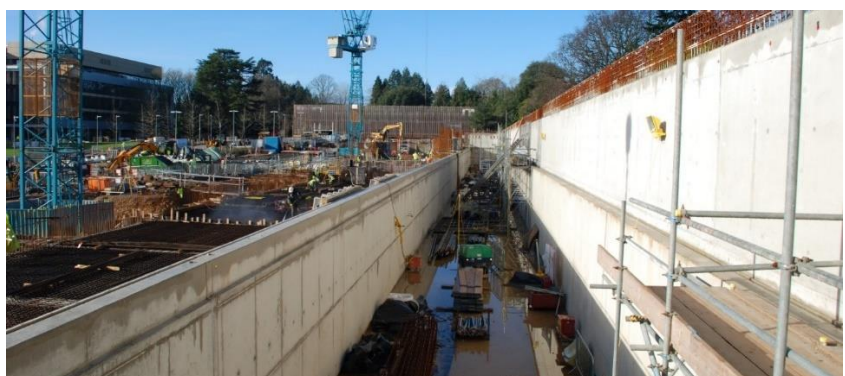

Figure 2 : Le site en Février 2014

---

[5] https://www.wates.co.uk/
[6] https://www.pdcm.pro/
[7] https://burofour.com/

En parallèle, les études pour les équipements scientifiques ont été lancées avec le soutien du bureau d'études Frazer-Nash Consultancy[8] (FNC) et de l'entreprise Rekan[9]. La vitesse requise pour le chariot et la longueur limitée du bassin ont rapidement orienté le choix technique vers un chariot à câbles, permettant d'atteindre des accélérations plus élevées par rapport à une solution plus classique de chariot avec moteurs électriques à bord, en s'affranchissant des limites de coefficient de friction roues/rails et en allégeant le chariot. Afin de gagner en précision, il a été également décidé d'une propulsion à 2 winches et 2 câbles.

L'entreprise écossaise Penman Ltd.[10] a gagné en 2014 les appels d'offres pour le chariot – sans le système de propulsion qui devait pour des raisons budgétaires faire l'objet d'un appel d'offres séparé – et pour une plage d'amortissement latérale passive automatisée. Les livraisons étaient initialement prévues en 2015.

De la même façon, l'entreprise HR Wallingford[11], basée à Oxford, a remporté l'appel d'offres pour le générateur de houle, avec une livraison également prévue en 2015. Malgré leur proposition de plage d'amortissement passive, il a été décidé de la concevoir en interne et de sous-traiter sa fabrication à une entreprise locale, estimant que les coûts seraient plus bas pour une performance similaire.

Le bâtiment a été livré à l'université au printemps 2015, avec environ 6 mois de retard. Le générateur de houle, la plage d'amortissement et le chariot livrés peu après et le bassin rempli à l'été 2015. Pendant le remplissage, des fuites par infiltration à travers le béton sont rapidement apparues en bas des parois verticales du bassin, côté sud (seul mur accessible au sous-sol du bâtiment). Ces fuites ont été traitées par injection de résine dans le béton pour les endroits les plus touchés. Les plus petites fuites se sont vite résorbées d'elles même grâce à l'action d'un composé chimique présent dans le béton utilisé. Les fuites potentielles au niveau des parois non accessibles n'ont pas été traitées.

Une première version de la plage d'amortissement latérale a été également installée dans le bassin à l'été 2015, mais le système n'était pas satisfaisant, Penman Ltd. ayant mal apprécié les contraintes et tolérances liées au trop petit renfoncement prévu et à la construction du béton, et ayant également choisi des matériaux trop lourds pour les panneaux. Il a ainsi été décidé de revoir la conception intégralement.

Après plusieurs mois de retard, l'appel d'offres pour le système de propulsion du chariot a été publié à l'automne 2015, et remporté par Penman Ltd., en collaboration avec une autre entreprise écossaise ACE Winches[12] et une entreprise du nord de l'Angleterre Controlhouse[13].

A l'automne 2016, Penman Ltd. a fait faillite et était déclaré en liquidation judiciaire. A cette date, la livraison et mise au point du système de propulsion du chariot avaient déjà été retardées à l'été 2017 malgré la fabrication terminée des winches. La deuxième version de la plage d'amortissement latérale était quant à elle en fin de conception.

Peu après la faillite de Penman Ltd., il parut évident de discuter avec ACE Winches pour la reprise du projet, les winches étant déjà fabriqués. Les processus de marchés publics et légaux ont abouti en mai 2018 au contrat signé entre l'université et ACE Winches, en partenariat cette fois avec la société anglaise Iconsys[14] pour la partie contrôle.

ACE Winches a également proposé un changement d'emplacement pour les winches, en plus de les installer sur un chariot autorisant la translation latérale, permettant ainsi de conserver l'angle de tire constant à 90° du tambour. A l'origine, il était prévu d'installer les winches à l'intérieur du bassin sur les murs extérieurs ; ACE Winches a préconisé de les installer sur le toit avec des poulies de renvoi pour les câbles. Ces deux modifications permettent d'allonger la longueur utile du bassin et donc le temps de mesure.

---

[8] https://www.fnc.co.uk/
[9] https://www.rekan.co.uk/
[10] http://www.penman.co.uk/
[11] https://www.hrwallingford.com/
[12] https://www.ace-winches.com/
[13] https://www.controlhouse.co.uk/
[14] https://www.iconsys.co.uk/

Après des discussions internes, la direction de l'université a refusé l'emplacement sur le toit à l'automne 2018, craignant des difficultés avec le permis de construire et une dégradation des relations avec le voisinage résidentiel.

En réponse, ACE Winches a suggéré de construire deux locaux dédiés aux winches à chaque extrémité du bâtiment. Cette solution a été validée par l'université et une demande de permis de construire déposée en décembre 2018, cependant refusée en avril 2019 par les autorités locales, notamment suite à des réserves posées par le voisinage. Après consultations avec le département juridique de l'université, il s'est avéré qu'en raison de leurs dimensions par rapport au reste du bâtiment, ces locaux ne nécessitaient pas de permis, mais qu'une déclaration en mairie suffisait. La construction a définitivement été entérinée en août 2019 et a commencé à l'automne suivant, en même temps que l'installation de l'électronique de puissance et des 6 kilomètres de câbles requis. Les winches ont été livrés en janvier 2020 et les toits posés peu après.

En parallèle, l'entreprise Scale Engineering[15], qui avait participé aux spécifications initiales de la plage d'amortissement latérale, a repris le travail sur ce système en 2019. En utilisant la dernière version produite par Penman Ltd. avant leur faillite, et en travaillant sur une maquette du mur sud pour résoudre les derniers problèmes, Scale Engineering a livré le système final en février 2020.

Quelques jours après le premier mouvement du chariot le 18 mars 2020, le gouvernement du Royaume-Uni déclarait le confinement de la population suite à l'épidémie de COVID-19, mettant en pause la mise au point du système de propulsion. En raison des disponibilités et du statut de personne à risque de personnels clés dans l'équipe de sous-traitants, le projet n'a pu repartir qu'en septembre 2020, mais a dû de nouveau s'arrêter de nouveau en janvier 2021 suite au second confinement au Royaume-Uni.

A l'été 2021, l'équipe technique était pour de bon de retour sur le site pour terminer la mise au point du système de propulsion. Les réglages fins du système ont été plus difficiles que prévu, et quelques soucis avec le système de freinage d'urgence et la conception du lien câbles-chariot ont encore causé quelques retards. Le système a été livré à l'université le 1$^{er}$ février 2022, avec 55 mois de retard sur le calendrier initial, et presque 10 ans après le début du projet.

## III - Description des équipements

### III - 1 Rails

A la fermeture du bassin de carènes de GKN en 2008, l'université a réussi à récupérer les bogies du chariot, les rails et leurs 440 platines de support et réglage. Si les bogies et les rails découpés de façon aléatoire ne convenaient pas, les platines ont été rénovées et utilisées pour le nouveau bassin. Elles ont été installées fin 2014. Les nouveaux rails, fournis par l'entreprise rapidrail[16] ont quant à eux été livrés début 2015 en tronçons de 10 m et soudés sur place.

La littérature (référence [8]), et notamment le travail effectué au bassin de l'Australian Maritime College (AMC) à Hobart en Australie (référence [9]) montrent que pour réaliser un alignement précis, les recommandations suivantes peuvent être appliquées :
- Une fois le bassin rempli, il faut laisser le béton se tasser avant d'aligner les rails. Les équipes de l'AMC ont mesuré des déformations jusqu'à 1,2 mm dans les 12 mois suivant le remplissage ;
- L'utilisation d'un laser ou d'un théodolite n'est pas recommandée à cause des gradients de température et d'humidité au-dessus du bassin ;
- Les rails doivent être alignés verticalement pour suivre la courbure de la Terre (à Southampton, pour un bassin de 138 m orienté est-ouest, la flèche de l'arc est d'environ 0,6 mm), avec une tolérance cible de ± 0,1 mm sur la longueur du bassin.

---

[15] https://scaleengineering.co.uk/
[16] https://www.rapidrail.co.uk/

L'alignement des rails du bassin de Boldrewood a débuté en janvier 2016, soit 8 mois environ après le remplissage du bassin. Les techniques suivantes, basées sur les travaux de l'AMC et améliorées, ont été utilisées :

- Pour la planéité de la surface supérieure (en roulis et tangage), un niveau à bulles de haute précision (0,2 mm/m) a été utilisé ;
- Les deux rails ont été réglés à la bonne distance entre eux à l'extrémité ouest du bassin avant l'alignement individuel. Le parallélisme a été ensuite vérifié après alignement sur toute la longueur du bassin en utilisant un télémètre laser de précision ;
- L'alignement longitudinal a été assuré en tendant un fil de pêche en Dyneema® de 150 m de long et 0,3 mm de diamètre entre deux petits treuils fixés à chaque extrémité du rail. Ces treuils, conçus et fabriqués en interne, permettent de tendre le fil avec une force d'environ 20 kg pour réduire sa flèche. Un microscope fixé sur un support posé sur le rail permet ensuite d'ajuster le rail latéralement en visant le fil (voir A et B sur la figure 3) ;
- L'alignement vertical est assuré grâce à une petite tranchée dans le béton, située à côté des platines et prévue dès la construction. Cette tranchée est remplie d'eau (environ 600 L). Deux barres en acier usinées en pointes très effilées et pointant vers le bas sont tenues par un support posé sur le rail. Réglables en hauteur, elles sont descendues jusqu'à « casser » la tension de surface de l'eau dans la tranchée. Le support est ensuite déplacé le long du bassin et chaque platine ajustée pour répéter le réglage obtenu précédemment (voir C et D sur la figure 3). Il faut vérifier le réglage cible régulièrement à cause de l'évaporation de l'eau dans la tranchée.

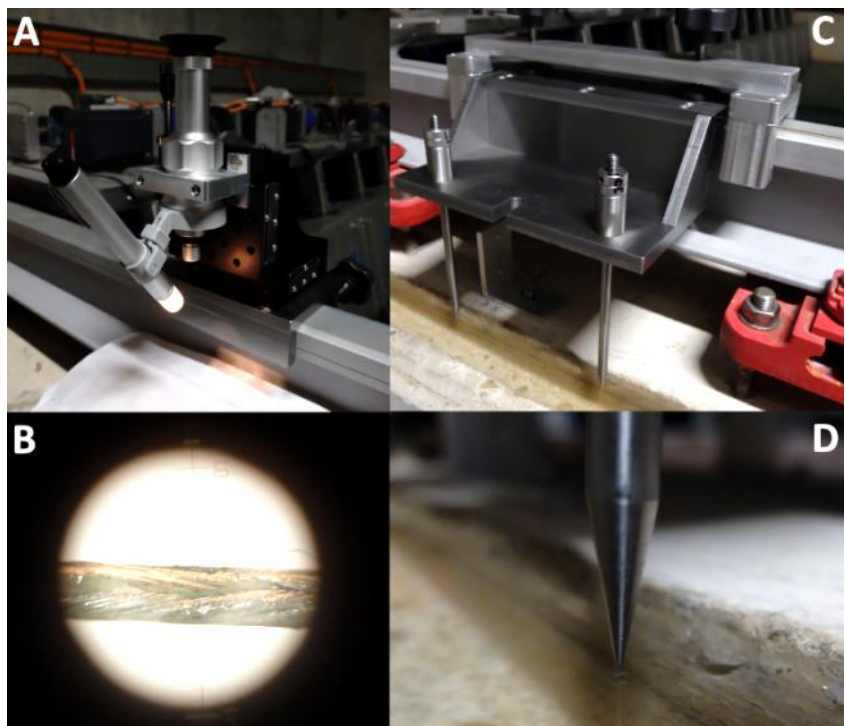

Figure 3 : Alignement des rails

Ces méthodes sont très précises, mais s'appuient sur la vision humaine. L'alignement des rails doit donc être réalisé par une seule personne. Le technicien du bassin a passé environ 3 semaines pour aligner chaque rail, et le rail nord n'ayant pas d'accès piéton, il a fallu l'aligner à partir d'un ponton flottant.

L'alignement vertical du rail sud a été vérifié de façon ponctuelle (une vingtaine de points sur la longueur) et régulière après 2016, et n'a pas montré de mouvement perceptible dans le béton.

Une fois le bassin livré, une vérification complète de l'alignement vertical des 2 rails a pu être réalisée en avril 2023 (référence [10]), en utilisant deux sondes à houle à ultrasons fixées de chaque côté du chariot et en faisant avancer le chariot à très basse vitesse.

Les résultats (Figure 4) montrent que le rail sud est presque parfaitement aligné dans la partie centrale de mesure et présente des défauts d'alignement aux extrémités, conséquences de dépôts de rouille, eux-mêmes causés par la présence des grilles de ventilation et par l'action moins efficace des brosses du chariot. L'alignement du rail nord est moins bon, avec des écarts entre -1,0 mm et +0,5 mm dans la partie centrale. Il est admis que ces résultats sont la conséquence directe de l'alignement réalisé à partir d'un ponton flottant. Même si la tolérance ciblée n'est pas atteinte, l'alignement des rails du bassin de Boldrewood est en moyenne satisfaisant, avec un écart moyen de 0,26 mm dans la partie centrale où les mesures sont faites.

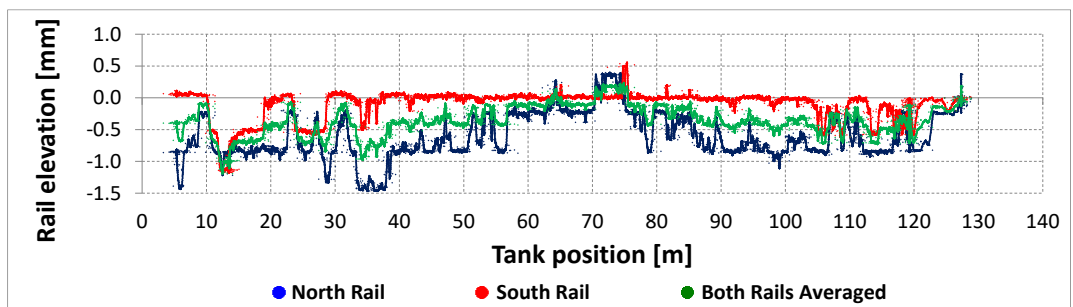

Figure 4 : Vérification de l'alignement vertical des rails

### III - 2 Traitement de l'eau

Le bassin de carènes de Boldrewood a été conçu sans fenêtres pour éviter la présence de courants et couches thermiques dans l'eau, mais aussi la croissance d'algues microscopiques. Tous ces éléments ont précédemment posé problème dans d'autres bassins.

Le bassin a été rempli avec 2900 m$^3$ d'eau douce en juin 2015, livrée par camions sur une période de 3 semaines. Peu après, à l'automne 2015, une « croûte » blanche s'était formée à la surface de l'eau. Après analyse, il s'agissait d'une forme de bactérie inconnue, mais non dangereuse pour les humains.

Une visite au bassin de traction de Centrale Nantes[17], de taille très proche, confirmait que le système de filtration installé était très sous-dimensionné. Avant la fin 2016, le système était entièrement démonté et remplacé par :

- Une pompe plus puissante atteignant un débit de 35 m$^3$/h ;
- Des tuyaux de diamètre 100 mm (50 mm pour les originaux) ;
- 2 grands filtres à sable ;
- Une lampe à ultraviolets ;
- Un système automatique d'injection de chlore.

Le système, prévu pour fonctionner uniquement la nuit, traite l'entièreté du volume du bassin en une semaine, et maintient un niveau de chlore à environ 1 ppm, évitant tout nouveau problème depuis son installation.

### III - 3 Le chariot

Le concept du chariot a été développé en collaboration avec FNC et consiste en une structure en anneau fabriquée en tôle d'aluminium, ce qui fournit la rigidité nécessaire sans treillis.

Le chariot a été conçu et livré en 4 sections qui ont été assemblées sur les rails (Figure 5) en utilisant 8 tiges filetées M24. Il a été pensé pour faciliter les essais, avec un plan ouvert et des accès faciles. Une plateforme motorisée permet l'accès à la maquette, une poutre de levage embarquée de 500 kg permet la manutention d'équipements partout dans le bassin et enfin des planchers amovibles permettent un accès facile sous le chariot, notamment pour l'installation du matériel photo et vidéo. Afin de maximiser le type d'essais réalisables, le chariot peut être utilisé dans les deux directions.

Le chariot est équipé de 4 bogies, chacun avec 2 roues principales en acier et 4 roues latérales en plastique pour empêcher les mouvements latéraux. Les bogies sud (côté corridor) sont équipés de protections en plexiglas sur mesure pour éviter l'accès au mécanisme et pour dévier tout objet présent sur le rail. Enfin, chaque coin du chariot est équipé avec une brosse métallique pour éviter l'accumulation de rouille sur les rails.

---

[17] https://lheea.ec-nantes.fr/moyens-dessais/bassins-de-genie-oceanique/bassin-de-traction

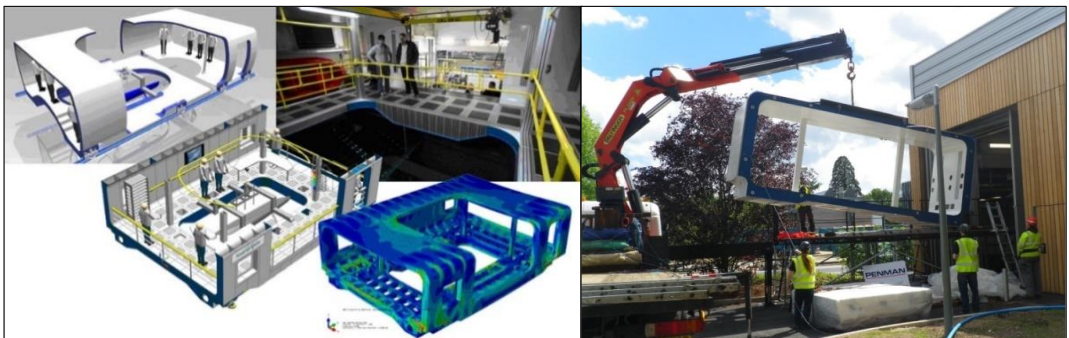

Figure 5 : Concept du chariot et livraison en sections

La propulsion du chariot est assurée par deux winches reliés au chariot par deux câbles de Dyneema© de 150 m de long et de diamètre 14 mm, et ayant une charge de rupture de 175 kN. Les winches sont entraînés par des moteurs électriques de 315 kW chacun et sont capables d'exercer une traction maximale de 35 kN, ce qui donne un facteur de sécurité de 5 pour les câbles.

Les winches sont montés sur des chariots transverses entraînés par des vis sans fin et asservis à la vitesse de rotation des tambours afin d'avoir une direction de traction dans l'axe longitudinal du bassin. Les points de tire, placés en haut du chariot, sont plus hauts que les winches. Il y a donc de chaque côté du bassin une poulie pour le renvoi d'angle des câbles à 90° vers le bas. Ces poulies sont instrumentées pour mesurer la tension dans chaque câble, données utilisées comme entrées par le système d'asservissement.

La position du chariot est mesurée à chaque instant par 3 lasers : 2 sont placés sur chaque face du chariot et un autre à l'extrémité ouest du bassin (précision ± 2,5 mm).

Le chariot peut être contrôlé à partir de 2 pupitres, un placé à bord et l'autre placé sur le bord du bassin, permettant de réaliser des essais sans personnel embarqué. Une console portable (Walk Around Box ou WAB) permet de piloter le chariot à partir des locaux des winches lors d'opérations de maintenance.

Si des chariots à propulsion par câble ont existé par le passé (par exemple le bassin de GKN), le système de propulsion du bassin de Boldrewood est à priori le premier à 2 winches et câbles (Figure 6).

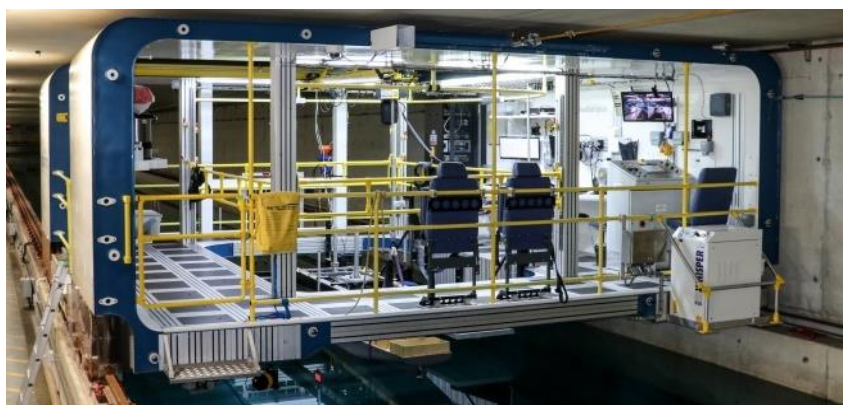

Figure 6 : Le chariot assemblé et opérationnel

Le choix d'un système à 2 winches, s'il est censé améliorer la qualité de la propulsion en comparaison d'un système plus classique, complexifie la partie contrôle. En effet, les 2 winches doivent être précisément asservis en temps réel en fonction des données mesurées (position du chariot, tension dans chaque câble, vitesse de rotation des winches) et des données calculées (vitesse du chariot, élongation des câbles).

L'élongation des câbles et les variations de longueur utile de chaque câble avec le déplacement du chariot sont les deux paramètres qui créent une relation non-linéaire entre la vitesse de rotation des moteurs et la tension dans chaque câble. Cette relation est résolue dans une boucle logicielle d'une durée de 1 ms. Le système de contrôle est de la gamme Siemens© Sinamics S120.

Afin d'assurer la sécurité des utilisateurs, et en se basant sur des recherches sur la chute de personnes dans les transports (références [11] à [13]), 3 modes d'opérations ont été définis en fonction de l'accélération du chariot, qui augmente progressivement avec la vitesse cible (Tableau 1) :
- Le mode I correspond à des accélérations inférieures à 0,10.g : les passagers peuvent rester debout ;
- Le mode II correspond à des accélérations entre 0,10.g et 0,20.g : les passagers doivent être assis ;
- Enfin, le mode III correspond aux accélérations supérieures à 0,20.g : les passagers doivent être assis et porter leur ceinture de sécurité.

| Speed m/s | Mode | Acc. m/s² | Acc. .g | Max. no passengers | Restrictions |
|---|---|---|---|---|---|
| 1 | I | 0.20 | 0.02 | 10 | Passengers can stand up. It is recommended to hold a safety rail. |
| 2 | I | 0.41 | 0.04 | 10 | Passengers can stand up. It is recommended to hold a safety rail. |
| 3 | I | 0.61 | 0.06 | 10 | Passengers can stand up. It is recommended to hold a safety rail. |
| 4 | I | 0.82 | 0.08 | 10 | Passengers can stand up. It is recommended to hold a safety rail. |
| 5 | I | 1.02 | 0.10 | 10 | Passengers can stand up. It is recommended to hold a safety rail. |
| 6 | II | 1.23 | 0.13 | 4 | Passengers to be seated (in a seat or on the floor). |
| 7 | II | 1.43 | 0.15 | 4 | Passengers to be seated (in a seat or on the floor). |
| 8 | II | 1.63 | 0.17 | 4 | Passengers to be seated (in a seat or on the floor). |
| 9 | II | 1.84 | 0.19 | 4 | Passengers to be seated (in a seat or on the floor). |
| >9 | III | 2.04 | 0.21 |  | Passengers seated and belted. |

Tableau 1 : Modes d'opérations

Le chariot est équipé de 3 systèmes de freinage :
- Le freinage normal est assuré par des freins à disques à air comprimé sur chaque winch. La vitesse de rotation du winch arrière est réduite, augmentant la tension dans le câble arrière et réduisant la vitesse du chariot. En même temps, la vitesse de rotation du winch avant est également réduite pour réduire la tension dans le câble avant. Ce freinage est automatique à la fin de chaque run et peut également être déclenché par le conducteur sur la console ou d'autres personnes en déclenchant un arrêt d'urgence (sur le chariot ou dans la halle d'essais). Ce mode de freinage est conçu afin que la décélération reste dans les limites prévues du run ;
- Si le freinage normal ne s'active pas dans le temps imparti, un freinage de secours est automatiquement déclenché, via une routine logicielle et des éléments électroniques différents certifiés pour les opérations de sécurité. Le winch arrière freine le chariot de la même façon que pour le freinage normal mais le winch avant est mis en roue libre. Le redémarrage du système requiert une remise en tension des câbles et dure donc plus longtemps ;
- Enfin, le dernier mode de freinage, le « crash-stop » ou arrêt anticollision empêche le chariot de percuter le mur ou le générateur de houle si les deux premiers modes de freinage ne fonctionnent pas et que le chariot dépasse les limites fixées en position et vitesse, si un des câbles se rompt ou si la position du chariot est perdue. Ce freinage est indépendant des winches et est obtenu par 8 plaquettes de frein actionnées par un compresseur à air en mode « sûr en cas de défaillance » situé sur le chariot (Figure 7). Les deux winches sont mis en roue libre, libérant la tension des câbles. Ce mode de freinage est plus violent que les deux premiers, c'est pourquoi il est entièrement automatique et ne peut être déclenché manuellement. Les plaquettes de frein ont une durée de vie limitée et doivent être vérifiées, voire changées après chaque arrêt.

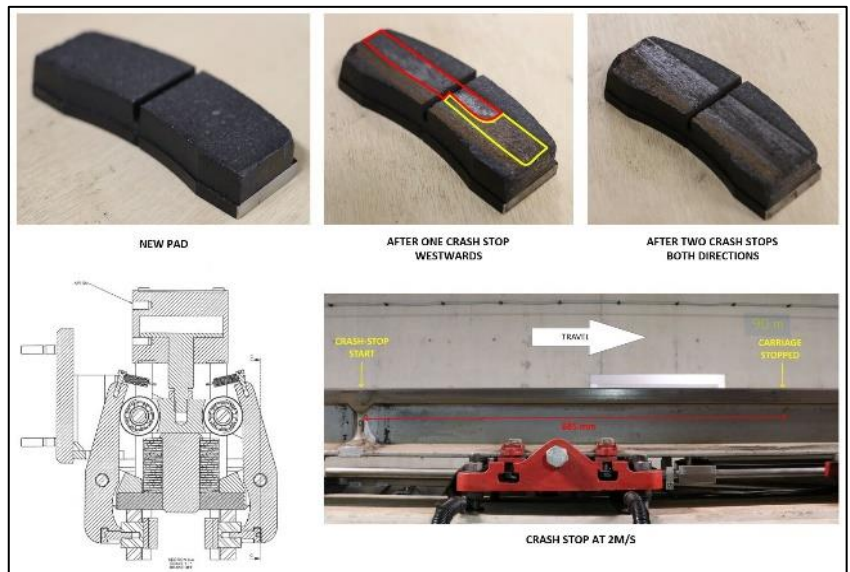
Figure 7 : Mécanisme du freinage « crash-stop »

III - 4 Le générateur de houle et les plages d'amortissement

Le générateur de houle HR Wallingford est constitué de 12 volets indépendants et est capable de générer des houles régulières jusqu'à 0,7 m de hauteur et des houles irrégulières de hauteur significative jusqu'à 0,35 m (Figure 8). Il est également possible de générer des vagues obliques, même si cette fonctionnalité a ses limites dans un bassin de carènes.

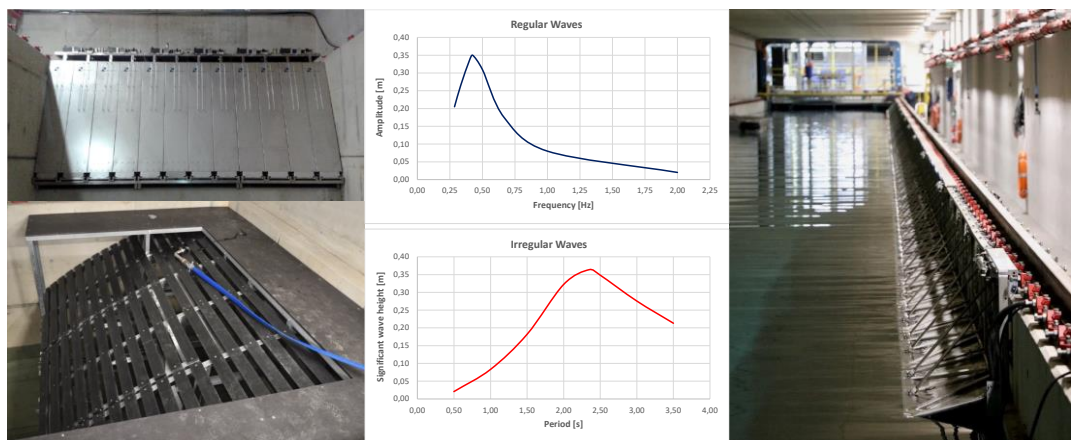
Figure 8 : Le générateur de houle et les plages d'amortissement

Une plage d'amortissement passive a été installée à l'autre extrémité du bassin. Ce choix de concevoir cette plage en interne a été rétrospectivement une erreur, les coûts de la conception et de la gestion de la fabrication ayant au final été supérieurs, et les performances de la plage n'ayant pas été satisfaisantes. Des corrections ont depuis été apportées afin d'améliorer cet équipement.

Une plage d'amortissement latérale passive mais automatisée a également été installée sur le mur sud du bassin, répliquant le système auparavant présent dans le bassin de GKN. Un total de 60 panneaux de 2 m de long chacun sont fixés au mur et opérés par groupes de 5 par 12 actionneurs linéaires synchronisés. Les panneaux se déploient de 100° par rapport à la verticale et sont en position basse immergés d'environ 90 mm ou 20% de leur hauteur. Le déploiement peut être déclenché à partir de la console du chariot ou de la console au bord du bassin et prends environ 10 secondes. Lors d'essais en eau calme, la plage est déployée en permanence, amortissant les vagues crées par la maquette. Lors d'essais de tenue à la mer, elle est remontée pendant les runs et déployée à la fin de chaque essai. Cette plage latérale est très efficace et permet des gains de temps conséquents par rapport à un système plus classique comme des lignes de piscine.

III - 5 Equipements de mesure

La balance de traction du bassin de Boldrewood a été conçu en interne par WUMTIA, dont c'est une spécialité. Elle consiste en une cage déformable dont les plaques « ressorts » sont interchangeables pour ajuster la gamme de mesure et la précision. Les forces de résistance à l'avancement et de dérive sont obtenues à partir de mesures des déformations de cette cage par capteurs de déplacement LVDT (Linear Variable Differential Transformer). La balance peut comprendre entre 1 et 3 pylônes de traction libres en pilonnement suivant la taille de la maquette. Elle permet également l'ajustement des angles de dérive et de gîte ainsi que la mesure du tangage, du moment de lacet et du moment de redressement suivant les configurations. Un logiciel d'acquisition sur mesure codé en LabVIEW a été sous-traité à l'entreprise locale SSDC[18], incluant une fonctionnalité de sauvegarde automatique et accès aux données à distance pour les utilisateurs sur les serveurs de l'université.

Deux systèmes de mesure de mouvements par trajectométrie ont été fournis par la société Qualisys[19], un système aérien (8 caméras Oqus et 2 caméras Miqus) et un système sous-marin (6 caméras Oqus). Les 2 systèmes peuvent opérer indépendamment ou être couplés et ont été utilisés pour de nombreux projets d'enseignement, de recherche ou commerciaux (Figure 9 et référence [14]), dans le bassin de Boldrewood mais également pour des projets de recherche sur la performance de nageurs dans des piscines.

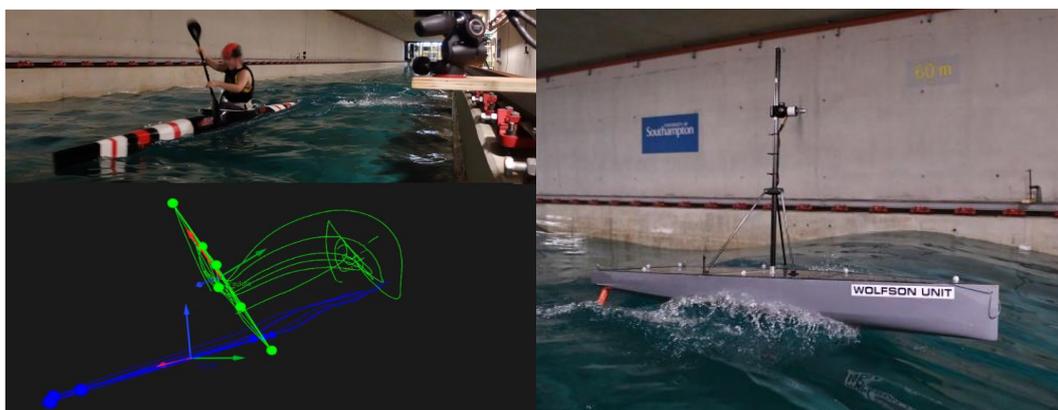

Figure 9 : Exemples d'utilisation des systèmes de trajectométrie (références [15] et [16])

Un jeu de 4 sondes à houle acoustiques de marque General Acoustics[20] a été acquis pour le bassin. Ces sondes peuvent être combinées en groupes de 2 ou 3 pour un seul canal de mesure afin de s'affranchir de problèmes bien connus de mesure de vagues cambrées par ce type d'équipement.

Un système de vélocimétrie par images de particules sous-marin a été fourni par l'entreprise LaVision[21]. Le système d'ensemencement installée sur le chariot a été conçu en interne en se basant et en améliorant le système de l'université TU Delft[22] aux Pays-Bas :
- Les particules de plastique sont pré-mixées dans un seau de 40 L ;
- Ce pré-mélange est ensuite pompé dans un circuit d'eau elle-même pompée du bassin ;
- Le tout est injecté dans la partie centrale du bassin durant le retour à la position de départ en utilisant un « râteau », qui est ensuite sorti de l'eau pendant les mesures.

Cet équipement a été mis au point et validé (références [17] et [18]) et peut également servir pour des mesures utilisant la corrélation numérique d'images (référence [19]).

---

[18] https://ssdc.co.uk/
[19] https://www.qualisys.com/
[20] https://www.generalacoustics.com/
[21] https://www.lavision.de/fr/applications/fluid-mechanics/underwater-piv/index.php
[22] https://www.tudelft.nl/me/over/afdelingen/maritime-and-transport-technology/research/ship-hydromechanics/facilities/towing-tank-no-1

### III - 6 Suivi environnemental

Afin d'assurer le suivi environnemental dans le bassin, et en réaction aux parfois mauvaises expériences vécues dans d'autres bassins, les capteurs suivants ont été installés :
- 6 capteurs de température de l'eau du bassin, situés à 3 positions longitudinales et 2 profondeurs ;
- 2 capteurs de température et d'humidité de l'air, situés aux deux extrémités de la pièce ;
- 1 capteur de pression de l'air, situé au milieu de la pièce ;
- 1 capteur de profondeur d'eau (par mesure de pression) situé juste devant le générateur de houle.

Un logiciel d'enregistrement de ces données a été développé par la société SSDC en LabVIEW, et une page intranet créée pour accéder à l'historique depuis le réseau de l'université.

Le suivi de la température de l'eau au cours des années a permis de s'assurer que le bassin ne présente pas de phénomène de stratification thermique, problématique entre autres pour des essais de voiliers. L'évolution de la température de l'eau au cours des saisons est quasi-identique aux différents endroits du bassin (0,2% de différence en moyenne) et la température moyenne a une amplitude entre 5 et 6°C sur une année, et entre 0,5 et 1,0 °C sur un mois. Ces faibles variations permettent de réaliser des essais de tous types dans d'excellentes conditions environnementales en éliminant les incertitudes présentes dans certains bassins.

## IV - Mise en service et validations

### IV - 1 Vitesse maximale du chariot

En raison d'un manque de budget et de temps à la fin de la mise en service du système de propulsion du chariot, la vitesse maximum a été réduite à 10 m/s dans la direction « normale » est-ouest et à 8 m/s dans la direction ouest-est. Au vu de la longueur du bassin, cette réduction n'a qu'une faible incidence. La différence de vitesse entre les deux directions n'a pas été expliquée par ACE Winches et Iconsys et pourrait être due à des petites différences géométriques entre les deux winches et leurs « accessoires ».

### IV - 2 Performances du chariot

A la réception finale du bassin en 2022, des essais de performance ont été réalisés pour le chariot et son système de propulsion (référence [20]). Les essais (Figure 10) ont consisté à réaliser des runs sans maquette avec une seule personne à bord dans les deux directions, et en augmentant progressivement la vitesse. Ils ont permis les observations suivantes :
- La différence entre la vitesse moyenne mesurée et la vitesse cible est nulle à basse vitesse et augmente avec la vitesse du chariot, jusqu'à atteindre 12 à 13 mm/s suivant la direction de déplacement (la vitesse mesurée étant toujours inférieure à la vitesse cible). Ces différences peuvent être expliquées par un léger manque de précision dans la mesure de la circonférence des deux winches ;
- L'« overshoot », ou période de stabilisation entre la phase d'accélération et la vitesse constante, est relativement constant autour de 2 s ;
- L'amplitude du bruit de la vitesse mesurée durant la phase de vitesse constante varie entre 25 mm/s et 50 mm/s, en augmentant avec la vitesse du chariot, pour des vitesses jusqu'à 8 m/s. Pour la direction est-ouest, cette amplitude augmente ensuite jusqu'à atteindre 150 mm/s à une vitesse de 10 m/s ;
- Le temps de mesure utile diminue de manière exponentielle avec la vitesse. Il est de 45,4 s à 2 m/s, 18,5 s à 4 m/s, 9,8 s à 6 m/s, 4,4 s à 8 m/s et 1,8 s à 10 m/s pour la direction est-ouest.

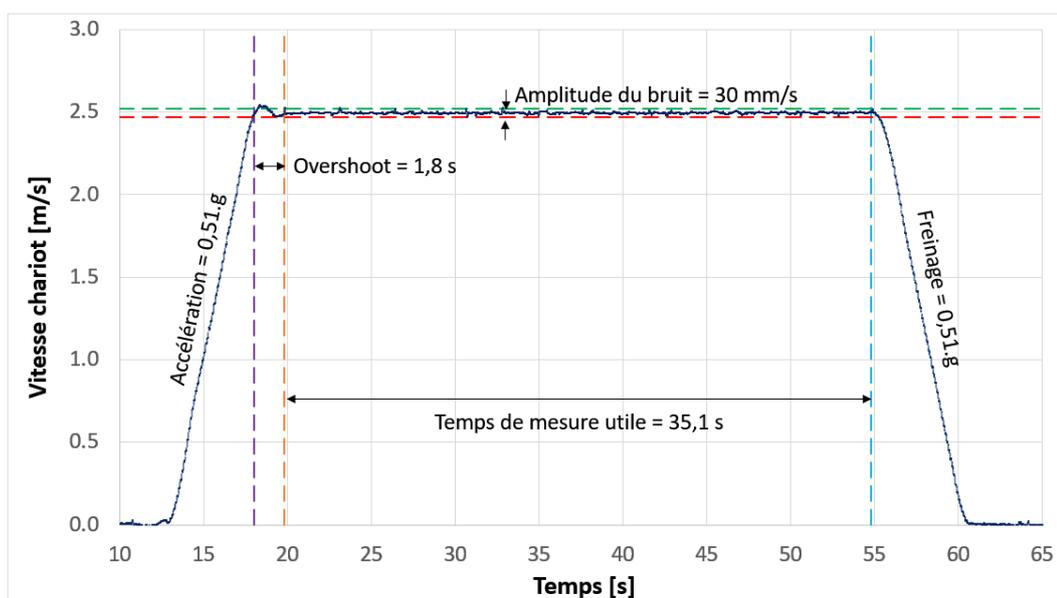

Figure 10 : Exemple de run de validation, vitesse 2,5 m/s, direction est-ouest

Ces performances peuvent être comparées à celles du bassin de traction de Centrale Nantes, pour lequel la même analyse a été réalisée (référence [21]). Ce bassin est de dimensions très proches et a été mis en service au début des années 2000. Le temps de mesure utile en fonction de la vitesse est très proche entre les deux bassins, ce qui est attendu au vu de leurs longueurs quasi-équivalentes (138 m pour Boldrewood et 140 m pour Centrale Nantes). La capacité du chariot à atteindre la vitesse cible est également similaire entre les deux bassins, tout comme l'amplitude du bruit de la vitesse mesurée. En revanche, la durée de l' « overshoot » est inférieure à Boldrewood, la stabilisation intervenant entre 1 s et 3 s plus rapidement.

Ces mesures prouvent que les performances du chariot du bassin de Boldrewood sont bonnes et suffisantes pour réaliser des essais en respectant les procédures de l'ITTC (référence [22]). La précision de la vitesse atteinte pourrait être améliorée par mesure précise des winches et/ou application d'une fonction de correction dans le logiciel de contrôle. D'après les experts d'Iconsys, l' « overshoot » et le bruit pourraient également être réduits, mais il faudrait pour ça pouvoir disposer de plusieurs semaines de réglages fins. A cette date, ceci n'a pu être fait par manque de temps et de budget.

IV - 3 Validation de la mesure de résistance

Les essais de validation du bassin ont eu lieu en mars 2022. Ils ont consisté en des essais de résistance sur eau calme en utilisant une maquette de KCS (Kriso Container Ship), qui avait été utilisée comme benchmark pour le workshop de Tokyo en 2015 (référence [23]).

La maquette utilisée en 2015 étant trop grande pour le bassin (échelle 31,5994, longueur hors-tout 7,7 m), une nouvelle maquette plus petite a été fabriquée (échelle 60,9547, longueur hors-tout 4,0 m, figure 11). Les résultats de Tokyo ont été mis à l'échelle de la nouvelle maquette en utilisant la méthode ITTC 1957.

La comparaison des résultats à l'échelle « Boldrewood » montre une différence moyenne de 0,7%, entre -0,9% et +1,8%. Si la comparaison n'est pas parfaite à cause de la différence de maquette et d'échelle, ces résultats valident les équipements du bassin de Boldrewood.

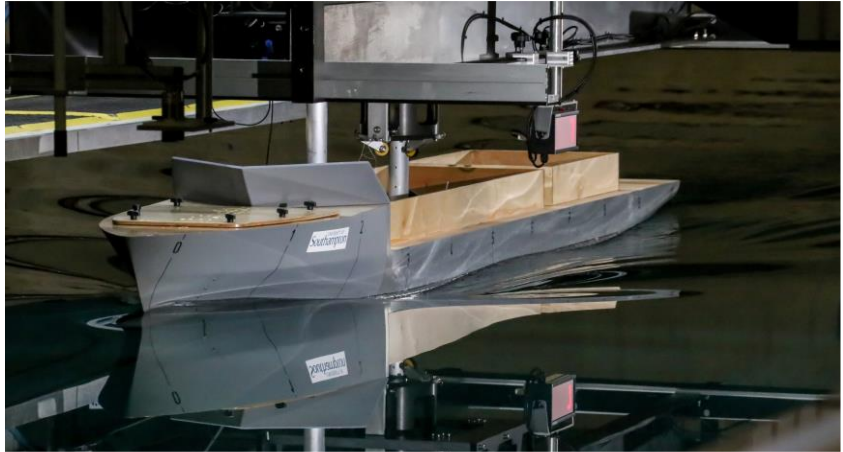
Figure 11 : Essais de validation avec la maquette KCS

## V- Leçons et conclusion

Rétrospectivement, les leçons à tirer des problèmes et délais rencontrés lors de la mise en service du bassin de carènes de Boldrewood sont :

- L'université a sous-traité la gestion des contrats liés à la construction du bâtiment et à la fourniture des équipements scientifiques à différentes sociétés elles-mêmes chapeautées par une autre société, elle-même sous responsabilité des services achats et technique qui consultaient les personnels scientifiques du bassin. Cette organisation a entraîné plusieurs problèmes :
  - Les responsabilités ont été diluées. Dans le cas de l'installation du premier système de filtration sous-dimensionné (III – 2), les délais de réaction ont été extrêmement longs, avec comme conséquence des coûts additionnels de traitements temporaires de l'eau. Par ailleurs, l'erreur initiale de dimensionnement, qui venait d'un cahier des charges peu précis de l'université, n'a pas été identifiée par ces entreprises ni par le fournisseur final, et l'université a dû payer l'intégralité de la facture du nouveau système ;
  - Ces entreprises externes ayant toutes des contrats avec l'université sans obligation de résultat ni pénalités de retard et à rémunération horaire, les coûts de gestion ont fortement augmenté au long du projet. La volonté de ces entreprises d'accélérer la résolution des nombreux problèmes rencontrés a été ressentie par beaucoup comme toute relative ;
  - Dans le cas du permis de construire des locaux accueillant les winches (II - 3), 4 mois ont tout d'abord été perdus à discuter entre toutes les parties et refuser la position proposée sur le toit, et ensuite 8 à 9 mois perdus à déposer et se voir refuser un permis de construire finalement non requis (ce qui n'a pas été identifié par ces entreprises spécialistes des projets de construction). Ces délais ont ensuite eu des conséquences très importantes en termes de retards avec l'apparition de l'épidémie de COVID-19 début 2020.
- La décision de construire le bassin ayant été prise très rapidement (II - 3), il n'y a eu que peu de consultations avec les fournisseurs et certains cahiers des charges ont été rédigés dans l'urgence :
  - Pour ces raisons et d'autres, ces fournisseurs n'ont pas souhaité répondre aux appels d'offres concernant le chariot et la plage d'amortissement latérale, entraînant un choix d'entreprise inexpérimentée dans le domaine de l'hydrodynamique expérimentale ;
  - Les dimensions du renfoncement de la plage latérale ont été décidées rapidement (II - 3). La profondeur de 60 mm retenue est tout juste suffisante et a causé des complications et donc des délais. On peut supposer qu'une profondeur de 100 mm aurait grandement facilité le projet ;
  - Le cahier des charges du chariot, et en particulier la vitesse maximale cible de 12 m/s, a entraîné un choix de système de propulsion innovant, voire inédit (II - 3). Ce choix a entraîné des délais de mise au point et des surcoûts qui n'ont pas permis d'en tirer le plein potentiel.
- La décision de refuser la proposition de plage d'amortissement de HR Wallingford après avoir estimé qu'elle était trop coûteuse et de réaliser la conception en interne avant de sous-traiter la fabrication et installation (II - 3) fut une erreur. La plage a en effet coûté plus cher et sa performance initiale n'était pas satisfaisante, même si des actions correctives ont depuis eu lieu.

Les retards du projet ont bien évidemment impacté le programme d'essais prévu. Malgré tout, le bassin a été utilisé pour des nombreuses activités d'enseignement, et quelques projets de recherche et industriels lors des 7 ans entre le remplissage et la livraison finale. Il a bien sûr fallu adapter les travaux pratiques et projets à un bassin sans chariot mobile, mais le bassin a néanmoins servi entre 50 et 125 jours par an lors de cette période.

Depuis début 2022, le bassin de carènes de Boldrewood est complétement opérationnel, performant et beaucoup utilisé. Pour l'année scolaire 2023-2024, il y a ainsi eu 150 jours d'activité, incluant 75 jours d'enseignement (travaux pratiques + projets pour un total de 235 étudiants impliqués), 42 jours de prestations commerciales et 33 jours de recherche, correspondant respectivement à 28%, 44% et 33% des revenus. Enfin, depuis 2015, le nombre total de visiteurs est estimé à 12000, allant de groupes de scolaires, de futurs étudiants à des chercheurs du monde entier en passant par des industriels de tous domaines.

## **Références**

[1] B. Malas, L. Creasey, D. Buckland et S. R. Turnock, Design, development and commissioning of the Boldrewood towing tank – A decade of endeavour, *International Journal of Maritime Engineering, 165 (3), A255-A271*, 2024.

[2] D. K. Brown, The way of a ship in the midst of the sea: the life and work of William Froude, *Periscope Publishing Ltd*, 2006.

[3] B. Deakin, Decades of development - 40 years history of the Wolfson Unit, *Ship & Boat International, RINA*, 2008.

[4] A. F. Molland, Resistance experiments on a systematic series of high-speed catamaran forms: variation of length-displacement ratio and breadth-draught ratio, *Transactions of The Royal Institution of Naval Architects, Issue 138, pp. 59-71*, 1996.

[5] A. F. Molland, P. A. Wilson, D. J. Taunton, S. Chandraprabha et P. A. Ghani, Resistance and wash measurements on a series of high-speed displacement monohull and catamaran forms in shallow water, *Transactions of The Royal Institution of Naval Architects Part A: International Journal of Maritime Engineering, Issue 146(2), pp. 19-38*, 2004.

[6] A. S. Bahaj, A. F. Molland, J. R. Chaplin et W. M. J. Betten, Power and thrust measurements of marine current turbines under various hydrodynamic flow conditions in a cavitation tunnel and a towing tank, *Renewable* Energy, Issue 32(3), pp. 407-426, 2007

[7] R. A. Cartwright, P. A. Wilson, A. F. Molland et D. J. Taunton, A low wash design for a river patrol craft with minimal environmental impact, *Transactions of The Royal Institution of Naval Architects Part A: International Journal of Maritime Engineering, Issue 150(A2), pp. 37-56*, 2008.

[8] P. Du Cane, High-speed small craft, 3rd Edition, *Temple Press Books, London*, 1964.

[9] A. Sprent et G. J. MacFarlane, The alignment of the Australian Maritime College towing tank, *Spatial Science Institute Biennial International Conference SSC2007, Hobart, Tasmania*, 2007.

[10] B. Malas, Boldrewood towing tank rail alignment check, *University of Southampton*, 2023.

[11] B. Malas, J. Banks, J. Cappelletto et B. Thornton, Applications of motion capture technology in a towing tank, *Advanced Model Measurement Technology for The Maritime Industry Conference, Rome*, 2019.

[12] F. Suva, Measuring coupled kayak and paddler motions within a towing tank, *Thèse de Mastère, University of Southampton*, 2017.

[13] E. Gauvain, The un-restrained sailing yacht model tests - A new approach and technology appropriate to modern sailing yacht seakeeping, *The 23rd Chesapeake sailing yacht symposium, Annapolis*, 2019.

[14] A. Lidtke et J. Banks, Underwater Particle Image Velocimetry (PIV) and dynamometry set up at the Boldrewood towing tank using BRIDGES thruster, *University of Southampton*, 2018.

[15] M. Gregory, J. Bowker, M. Kurt, J. Banks et S. Turnock, Underwater PIV benchmark at the University of Southampton Boldrewood Towing Tank, *National Wind Tunnel Facility*, 2023.

[16] H. Araujo Bento de Faria, Measurement of underwater deformation on simplified propeller blades using digital image correlation, *Thèse de Master, University of Southampton*, 2019.

[17] C. N. Abernethy, G. R. Plank et E. O. Sussman, Effects of deceleration and rate of deceleration on live seated human subjects, *U.S. Department of Transport report no. UMTA-MA-06-0048-77-3*, 1977.

[18] D. Martin et D. Litwhiler, An investigation of acceleration and jerk profiles of public transportation vehicles, *American Society for Engineering Education*, 2008.

[19] J. P. Powell et R. Palacin, Passenger stability within moving railway vehicles: limits on maximum longitudinal acceleration, *Urban Rail Transit 1, pp. 95–103*, 2015.

[20] B. Malas, Boldrewood towing tank carriage performance analysis, *University of Southampton*, 2022.

[21] B. Malas, Analyse de la performance du chariot du bassin de traction de Centrale Nantes, *LHEEA, Centrale Nantes*, 2024.

[22] ITTC, Resistance Test, *ITTC Guideline 7.5–02-02-01, Revision 5*, 2021.

[23] T. Hino, F. Stern, L. Larsson, M. Visonneau, N. Hirata et J. Kim, Numerical Ship Hydrodynamics – An assessment of the Tokyo 2015 workshop, *Springer. ISBN 978-3-030-47571-0*, 2020.